\documentclass[aps,prl,twocolumn,amsmath,amssymb, superscriptaddress,tightenlines]{revtex4}
\usepackage{graphicx}
\usepackage{epsfig}
\usepackage{makecell}
\usepackage{dcolumn}
\usepackage{bm}
\usepackage{amssymb}
\usepackage{bm}
\usepackage{amsmath, bm}
\usepackage{upgreek}
\usepackage{enumitem}  
\usepackage[colorlinks,citecolor=blue,linkcolor=blue,hyperindex]{hyperref}

\begin{document}

\title{Locally controlled arrested thermalization} 
\author{Ken K. W. Ma$^*$}
\affiliation{National High Magnetic Field Laboratory, Tallahassee, Florida 32310, USA}
\author{Hitesh J. Changlani}
\affiliation{National High Magnetic Field Laboratory, Tallahassee, Florida 32310, USA}
\affiliation{Department of Physics, Florida State University, Tallahassee, Florida 32306, USA}
\date{\today}

\begin{abstract}
The long-time dynamics of quantum systems, typically, but not always, results in a thermal steady state. The microscopic processes that lead to or circumvent this fate are of interest, since everyday experience tells us that not all spatial regions of a system heat up or cool down uniformly. This motivates the question: Under what conditions can one slow down or completely arrest thermalization locally? Is it possible to construct realistic Hamiltonians and initial states such that a local region is effectively insulated from the rest, or acts like a barrier between two or more regions? We answer this in the affirmative by outlining the conditions that govern the flow of energy and entropy between subsystems. Using these ideas we provide a representative example for how simple quantum few-body states can be used to engineer a ``thermal switch" between interacting regions.
\end{abstract}

\maketitle

\def\thefootnote{*}\footnotetext{Left academia}

\textit{Introduction--}
A central objective in nonequilibrium physics is to predict the dynamical behavior of isolated and open quantum systems in the long-time limit. Typically, the system will reach a thermal state that may be described by the eigenstate thermalization hypothesis (ETH)~\cite{Deutsch1991, Srednicki1994, Rigol2008}. However, recent developments have demonstrated examples where the ETH is violated~\cite{Altshuler2006, Huse2007, Huse2010, Huse2015, Abanin-RMP2019, Papic2021, Papic, Regnault2022, Moessner2023, Lukin2017, Turner2018, Lee_PRBR2020, McClarty_PRB2020, Lee_PRB2021}. This originates from the complete or partial breaking of ergodicity.

Here, we explore the possible impact on thermalization when the interaction between, or the initial state of, a \textit{few} spins in a local region of the system is altered, a scenario that can be realized in ultracold atoms with tunable interactions~\cite{Bloch2017, Kuklov2003, Lukin2003, Cirac2003, Altman2003, Ketterle2020, Ketterle2021, Chung2021, Ketterle2022}. The impact of such local perturbations can be dramatic---for example, one of us (in collaboration with others) recently found that a locally driven spin can be effectively decoupled from the rest of the system and remains athermal~\cite{RMelendrez_2022, BMukherjee_2023}. Alternately, one can locally heat and cool regions through local deformation of the Hamiltonian~\cite{Sondhi2019, Zaletel2021, Zoller2021, Ryu2021, Vishwanath2022}. It is thus useful to consider thermalization in different local regions of a system and their dependence on initial conditions (including inhomogeneous states). In addition to the conceptual question being interesting in its own right, such explorations may have practical implications as well. Preventing energy and entropy flow between different spatial regions translates to realizing a ``thermal switch" - it is ``on" (allows energy flow) or ``off" (completely stops energy flow) depending on which initial quantum state the system is prepared in. In this Letter we develop a framework for the conditions that are required to arrest thermalization in a local region. 

\begin{figure}
\includegraphics[width=\linewidth]{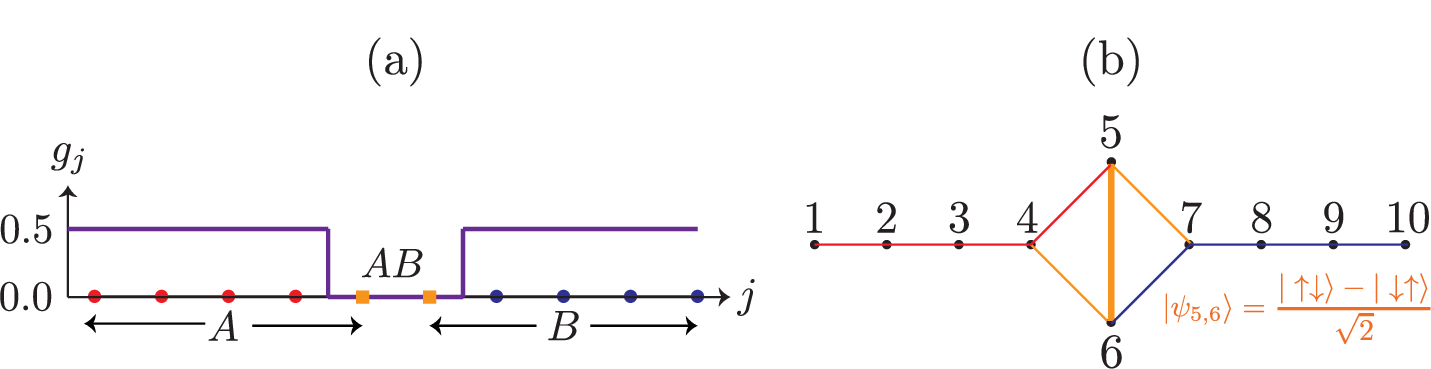}
\caption{Panel (a) shows the bipartite one-dimensional Ising chain under an external magnetic field, with $g_j$ being the field strength at site $j$. To achieve commuting subsystem Hamiltonians, the magnetic field at the boundary (i.e., the fifth and sixth spins) is set to zero. Panel (b) shows the Heisenberg chain with a thermal switch (formed by the fifth and sixth spins). Here, dots, squares (left panel), and bonds (right panel) in red, blue, and orange are in subsystems $A$, $B$, and their intervening region $AB$, respectively. Note that $AB$ has support in both $A$ and $B$.}
\label{fig:systems}
\end{figure}

Consider an isolated quantum many-body system that is divided into two regions $A$ and $B$ with strictly local Hamiltonians and an intervening region $AB$ which has support in $A$ and $B$. Our objective is to \textit {selectively} (not trivially) decouple these two regions, i.e., completely stop energy flow between them only for a certain (but fairly generic, and easy to realize) choice of initial conditions. The system is described by the Hamiltonian,
\begin{eqnarray}
H=H_A+ H_B+H_{AB}.
\end{eqnarray}
where $H_A$ and $H_B$ denote the Hamiltonians of region $A$ and $B$ respectively, and $H_{AB}$ is the coupling between the regions. Since the Hamiltonians considered are strictly local, $[H_A, H_B]=0$. We consider three distinct cases:
(1) $\left[H_A, H_{AB} \right] \neq 0$ and $\left[H_B, H_{AB} \right]\neq 0$; (2) $\left[H_A, H_{AB}\right]=0$ and $\left[H_B, H_{AB} \right]=0$; and (3) 
$\left[H_A, H_{AB} \right] = 0$ and $\left[H_B, H_{AB} \right] \neq 0$.

Case 1 is the most generic one, and we expect all the usual characteristics of ETH to be seen for \textit{most}, but not \textit{all} initial conditions. It will be our objective to show how a ``thermal switch" can be realized for this case. Cases 2 and 3, while distinct, will be shown to share many similarities in their behavior. We mention that aspects of heat flow and entropy production for commuting and non commuting subsystems have been previously addressed in the field of quantum thermodynamics~\cite{Kosloff2013, Pekola2015, Anders2016, Goold2016, Anders2017, Lostaglio2019, Deffner2019, Binder2019, Mitchison2019, Paternostro2021, Hossein-Nejad2015}, however, less attention has been devoted to a study of (1) physical observables beyond the energy and (2) the role of specific initial conditions in governing the long-time behavior.

\textit{Controlling thermalization in commuting subsystems--} Consider case 2, for example. $[H_A, H_{AB}]=0$ implies $[H_A,H]=0$; i.e. $H_A$ is a constant of motion. Even if $B$ is initially prepared to be ``hot" (say by initializing it in a random state), and $A$ ``cold" (i.e. by choosing an initial state with a low expectation value of energy), there will be no energy flow into subsystem $A$ even though it is coupled to $B$ via $H_{AB}$. How about the evolution of the quantum state itself and the local physical observables if the system is initially prepared in a separable state, $|\Psi(t=0)\rangle=|\psi\rangle_A\otimes|\varphi\rangle_B$?

The commuting conditions of case 2 imply $e^{-i(H_A+H_B+H_{AB})t}= 
e^{-iH_A t} e^{-i H_B t}e^{-i H_{AB}t}$. Hence,
\begin{eqnarray} \label{eq:evolution}
|\Psi(t)\rangle
=e^{-iH_A t}e^{-iH_B t}
e^{-iH_{AB}t}|\psi\rangle_A \otimes|\varphi\rangle_B.
\end{eqnarray}
By definition, $H_A$ and $H_B$ act on different regions. Meanwhile, the term $H_{AB}= \sum_j \mathcal{O}^j_A\otimes\mathcal{O}^j_B$ (we consider the case of only one term in the sum and hence drop the label $j$) acts on a small number of spins in the intervening region, for example, two spins in the Ising spin chain. The action of $e^{-iH_{AB}t}$ on the product state in Eq.~\eqref{eq:evolution} will typically generate entanglement between $A$ and $B$. However, $|\Psi(t)\rangle$ remains a product state \textit{if} $|\psi\rangle_A$ is a nondegenerate eigenstate of $H_A$, or if the initial state of the spins in $AB$ is an eigenstate of $\mathcal{O}_A$. In both cases, 
$\mathcal{O}_A|\psi\rangle_A=\alpha|\psi\rangle_A$. Hence,
\begin{eqnarray} \label{eq:HAB-act}
e^{-i\mathcal{O}_A\otimes\mathcal{O}_Bt}
|\psi\rangle_A\otimes |\varphi\rangle_B
=|\psi\rangle_A\otimes
\left(e^{-i\alpha \mathcal{O}_Bt}|\varphi\rangle_B\right).
\end{eqnarray} 
This leads to
\begin{eqnarray}
|\Psi(t)\rangle
&=e^{-iH_At}|\psi\rangle_A\otimes
\left(e^{-iH_Bt}e^{-i\alpha \mathcal{O}_Bt}|\varphi\rangle_B\right).
\end{eqnarray}
For an operator $O_A$ that only acts on subsystem $A$,
\begin{eqnarray}
\langle O_A(t)\rangle
=\left._A\langle\psi| e^{iH_At} O_A e^{-iH_At} |\psi\rangle_A.\right.
\end{eqnarray}
Therefore, any local measurement in $A$ will depend solely on the dynamics of $A$. In other words, regions $A$ and $B$ can thermalize in their own ways although they are coupled by $H_{AB}$. In particular, $\langle O_A(t)\rangle=\langle O_A(0)\rangle$ if $|\psi\rangle_A$ is an eigenstate of $H_A$ or if $O_A$ commutes with $H_A$. Meanwhile, an application of the ETH for subsystem $B$ suggests that $\langle O_B(t)\rangle\rightarrow\langle O_{B,\text{th}}\rangle$, where $\langle O_{B,\text{th}}\rangle$ is the value predicted from the thermal average.

The discussion above suggests a simple yet general recipe to control thermalization in different local regions in a quantum system. In quantum spin systems, the proposal can be realized by controlling the interaction between a small number of spins on the boundary between $A$ and $B$, which we elucidate with an example. 

\begin{figure}
\center
\includegraphics[width=\linewidth]{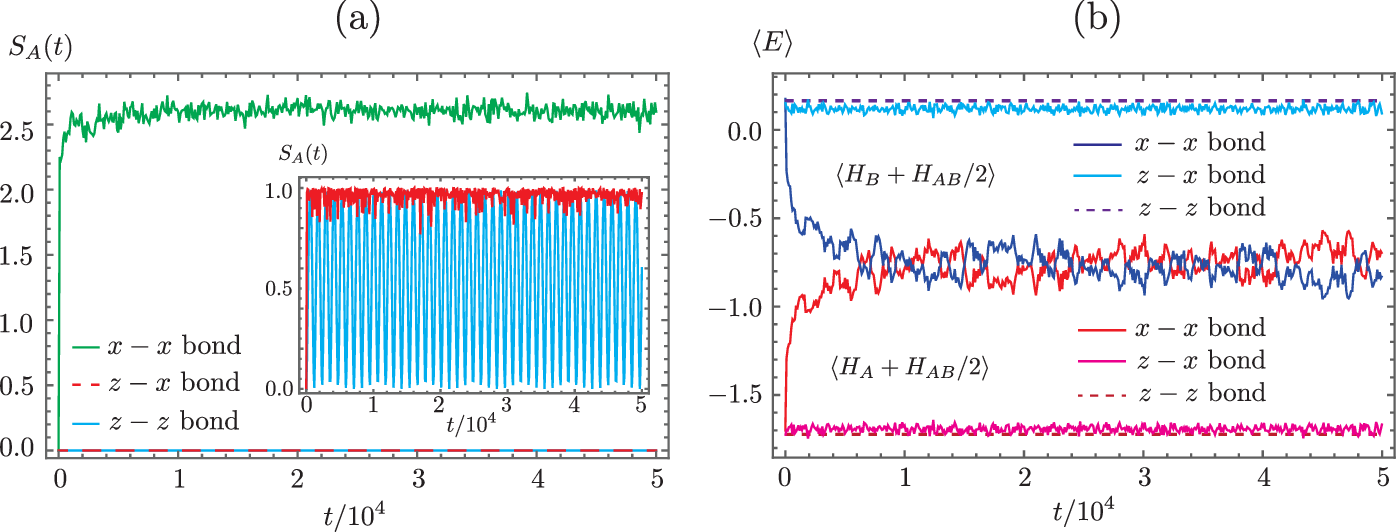}
\caption{Time evolution of the half-cut entanglement entropy between the two subsystems [(a)] for the tilted field Ising model described by Eqs.~(6) and (7), and the expectation values of energy for the subsystems [(b)]. The spin chain consists of ten spins. Each subsystem has five spins. Three different scenarios (the spin-spin interaction for the two spins at the boundary) are considered, which are explained by the legends in the figure. The initial states for all three cases are $|\uparrow\uparrow\uparrow\uparrow\uparrow\rangle_A\otimes|\text{ran}\rangle_B$, where $|\text{ran}\rangle_B$ stands for a random state for the five spins in subsystem $B$. The inset of (a) shows the half-cut EE of the system when $A$ is prepared in the initial state $|x\rangle^{\otimes 5}$, where $|x\rangle=(|\uparrow\rangle+|\downarrow\rangle)/\sqrt{2}$. }
\label{fig:SA-time-Ising}
\end{figure}

\begin{figure} 
\center
\includegraphics[width=\linewidth]{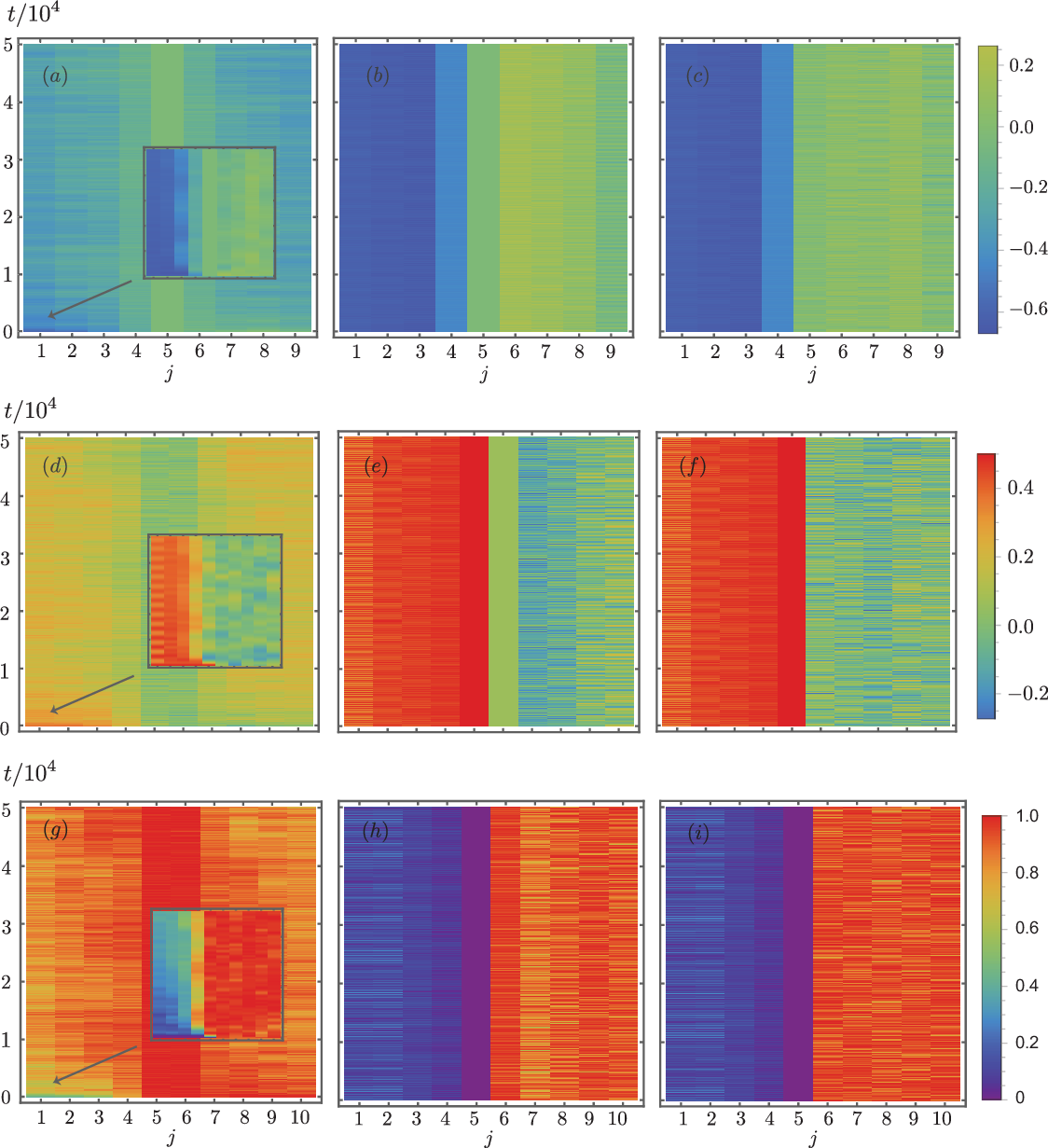}
\caption{Spatiotemporal profile of the estimated energy at each bond [(a)--(c)], spin magnetization $\langle S_j^z\rangle$ for all individual spins [(d)--(f)], and their corresponding von Neumann entropy [(g)--(i)] in the tilted field Ising model. Three different kinds of boundary are considered. For the left, middle, and right columns, the interaction terms are $H_{AB}=-S^x_5 S^x_6$, $H_{AB}=-S^z_5 S^z_6$, and $H_{AB}=-S^z_5 S^x_6$, respectively. In all cases, the initial state of the system is $|\uparrow\uparrow\uparrow\uparrow\uparrow\rangle_A\otimes|\text{ran}\rangle_B$. Here, we simulate the dynamics of the system for $0\leq t\leq 5\times 10^4$. The insets of the left column show the results in the short time interval $0\leq t \leq 100$. See the main text for more details.}
\label{fig:results-Ising}
\end{figure}

To illustrate our idea we consider a modification of the one-dimensional Ising spin chain in a tilted magnetic field, described by the Hamiltonian,
\begin{align} \label{eq:H2}
H_{A, B}
&=-J\sum_{i} S_i^z S_{i+1}^z
-g\sum_{j}
(\sin\theta S_j^x+\cos\theta S_j^z),
\\ 
H_{AB}
&=-J_{AB} S_{N_A}^{\sigma} S_{N_A+1}^{\sigma'}.
\end{align}
The tilted field Ising model was established to be non-integrable in Ref.~\cite{Lakshminarayan2007} (see also Ref.~\cite{Bastianello}). The indices $1\leq i,j \leq N_A-1$ for $H_A$. For $H_B$, $N_A+1 \leq i \leq N-1$ and $N_A+2\leq j\leq N$ where $N_A$ is the number of sites in $A$ and $N$ is the total number of sites. Here, we have set the magnetic field to be zero at sites $j=N_A$ and $j=N_A+1$. This allows us to engineer the interaction (i.e., by choosing $\sigma$ and $\sigma'$ in $H_{AB}$) such that different conditions for $[H_A, H_{AB}]$ and $[H_B, H_{AB}]$ can be realized.

As a proof of concept, we employ numerical exact diagonalization to study the dynamics of the system with ten spins, and fix $N_A=5$. We set the parameters $J=J_{AB}=1$, $g=0.5$, and $\theta=\pi/4$. The system is shown in the left panel of Fig.~\ref{fig:systems}. The initial state for subsystem $A$ is set to the ferromagnetic state $|\uparrow\uparrow\uparrow\uparrow\uparrow\rangle$, which is a low-lying energy state but not an eigenstate of $H_A$ due to the direction of the magnetic field. Meanwhile, subsystem $B$ is prepared in a random initial state. To quantify the average energy stored in subsystems $A$ and $B$, we calculate the expectation values $\langle H_A+H_{AB}/2\rangle$ and $\langle H_B+H_{AB}/2\rangle$ as a function of time. If no effective thermalization \textit{between} $A$ and $B$ occurs, the two expectation values should remain well separated from each other. Otherwise, they will eventually come close to each other.

We first set $H_{AB}=-S_5^x S_6^x$ to achieve case 1. The bipartite entanglement entropy (EE) between $A$ and $B$ is shown in Fig.~\ref{fig:SA-time-Ising}(a). The bipartite EE reaches the saturated value rather quickly, which indicates that the two subsystems are effectively entangled. From Fig.~\ref{fig:SA-time-Ising}(b), we also observe that $\langle H_A+H_{AB}/2\rangle$ and $\langle H_B+H_{AB}/2\rangle$ eventually come close to each other. This signifies the energy flow from $B$ to $A$ through the boundary, such that the two subsystems can thermalize. 

To demonstrate that genuine thermalization has been reached between the two subsystems, we obtain the energy stored at each bond, the spin magnetization $\langle S_j^z(t)\rangle$ for all individual spins, and their corresponding von Neumann entropy in base $2$ for $0\leq t\leq 5\times 10^4$. These are shown as spatiotemporal plots in Fig.~\ref{fig:results-Ising}(a), \ref{fig:results-Ising}(d), and \ref{fig:results-Ising}(g), respectively. The von Neumann entropy of a single spin is determined from the reduced density matrix obtained by partial tracing out all other nine spins. Since $H$ is not translationally invariant (no magnetic field acting on spins $5$ and $6$, and their spin-spin interaction is different from other spins), it is natural that the four spins close to and on the boundary behave differently from other spins. Meanwhile, the energy stored in bonds $1-3$ and $7-9$ are close to each other. Additionally, the spin magnetization and von Neumann entropy for those spins away from the boundary are close to each other. 

We then move to cases 2 and 3 by setting $H_{AB}=-S_5^z S_6^z$ and $H_{AB}=-S_5^z S_6^x$, respectively. Using procedures and definitions similar to the first case, we show the corresponding results for the present two cases in Fig.~\ref{fig:SA-time-Ising} and Fig.~\ref{fig:results-Ising}. In contrast to the first case, the bipartite EE stays at zero, and the average energy for the two subsystems remain separated from each other. The panels in the middle and right columns of Fig.~\ref{fig:results-Ising} further demonstrate that the boundary acts as a barrier, such that $A$ and $B$ evolve as if they are insulated from each other. These results provide examples of impeding thermalization in commuting subsystems.

What happens if the initial ferromagnetic state of $A$ is changed, but the initial state of $B$ remains unchanged? On one hand, $H_A$ remains a constant of motion in cases 2 and 3. On the other hand, the change in the initial state will impact the dynamics of subsystem $A$, and hence the local measurement of physical quantities there. To highlight this, we set the initial state of $A$ to $\left[\left(|\uparrow\rangle + |\downarrow \rangle\right)/\sqrt{2}\right]^{\otimes 5}$ in which case Eqs.~(3)--(5) no longer hold. Although not shown here, we find that the spatiotemporal profiles of the bond energy, $\langle S_j^z\rangle$, and the von Neumann entropy of the spins differ from those in Fig.~\ref{fig:results-Ising}. In the inset of Fig.~\ref{fig:SA-time-Ising}(a), we show the bipartite EE for cases 2 and 3. Instead of being a constant at zero, the EE now oscillates between zero and one in case 2, or increases from zero and saturates to one in case 3. This difference distinguishes between the two cases. Note that the entropy value of one (in base two) stems from the maximal entanglement between the two spins in the region $AB$. Thus, \textit{both} the Hamiltonian and initial states of the subsystems play a role in the dynamics and thermalization of the system.

\textit{Arrested thermalization for non-commuting subsystems--} We now address case 1 and ask whether it is possible to arrest thermalization at the local level when the commuting condition is not satisfied. Taking inspiration from work on global quantum many-body scars~\cite{Papic2021, Papic, Regnault2022, Moessner2023}, we show that this is indeed possible by constructing an explicit example. The general principle is to design a Hamiltonian $H_{AB}$ such that (1) the region $AB$ has a ``simple to prepare" eigenstate and (2) this eigenstate is unchanged by the time evolution in the rest of regions $A$ and $B$. We leverage the exact quantum degeneracy of certain locally frustrated motifs to accomplish this objective.

\begin{figure} 
\center
\includegraphics[width=\linewidth]{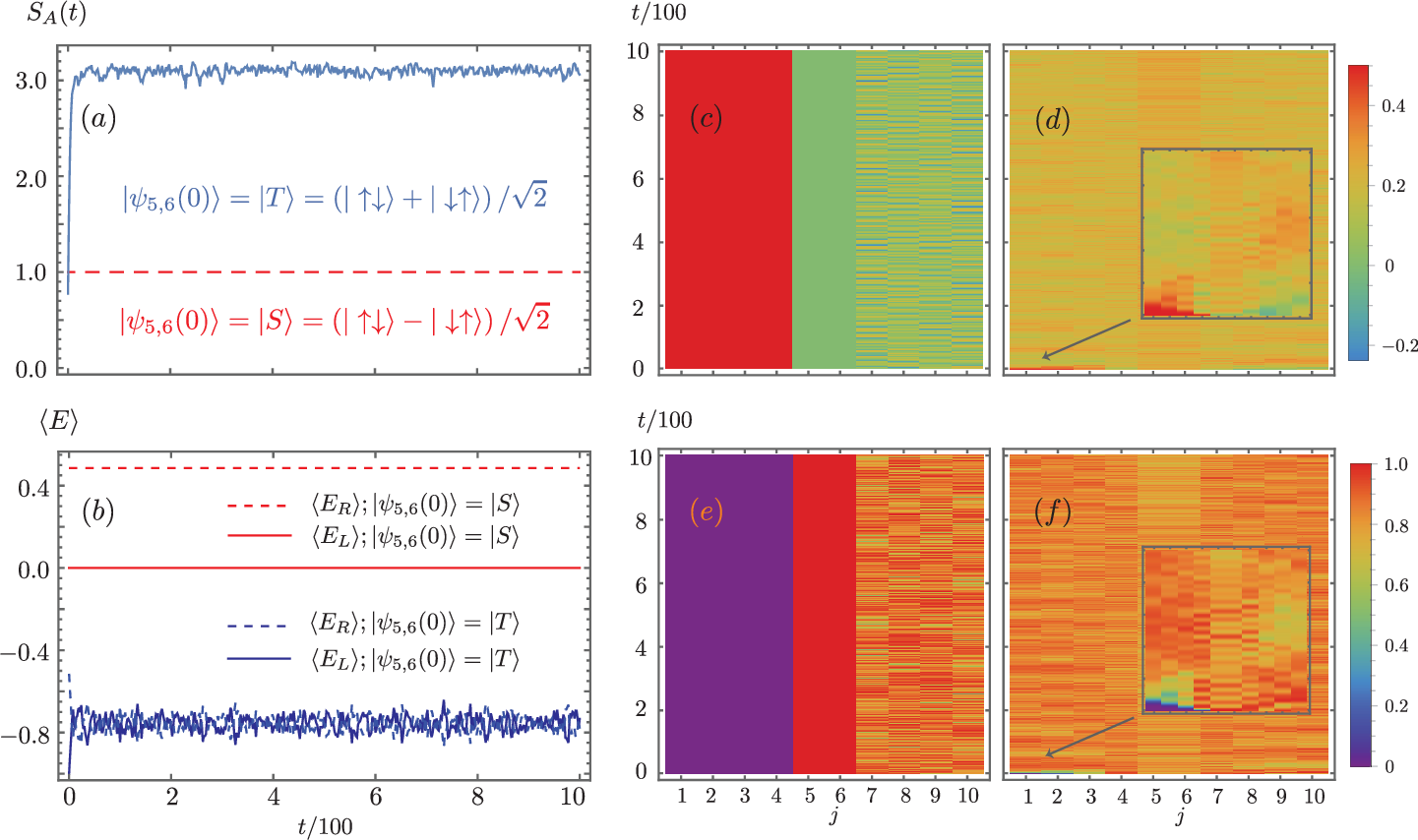}
\caption{Time evolution of the bipartite entanglement entropy with subsystem $A$ defined as the first five spins [(a)], and the expected energy for the left and right subsystems [(b)]. The left subsystem is defined as the spins $1-4$ and the ``left triangle" formed by spins $4,5,6$, whereas the right subsystem is defined as the ``right triangle" formed by spins $5,6,7$ and the spins $7-10$ (see Fig.~\ref{fig:systems}). (c) and (e) show respectively the time evolution of $\langle S_j^z(t)\rangle$ and the von Neumann entropy for every individual spin. Here, the initial state of the bond (spins $5$ and $6$) is set to the singlet state $|S\rangle=(|\uparrow\downarrow\rangle-|\downarrow\uparrow\rangle)/\sqrt{2}$. In (d) and (f), the corresponding results for a triplet initial state of the bond $|T\rangle=(|\uparrow\downarrow\rangle+|\downarrow\uparrow\rangle)/\sqrt{2}$ are illustrated. The insets show the short-time behavior from $t=0$ to $t=50$.} 
\label{fig:results-motif}
\end{figure}

Consider a lattice geometry that involves two triangles ``pasted together" as shown in Fig.~\ref{fig:systems}(b). As before, region $A$ consists of sites $1$ to $5$, region $B$ consists of sites $6$ to $10$, $H_A$ comprises bonds $(i,i+1)$ for $i=1-4$  and $H_B$ comprises bonds $(i,i+1)$ for $i=6-9$. $H_{AB}$ now consists of bonds $(4,6),(5,7)$ and $(5,6)$. The triangles share a common bond [here $(5,6)$], and each one of the unshared vertices connects to the remainder of the sites in either region $A$ or $B$. (Periodic arrangements of such motifs were previously considered in Ref.~\cite{McClarty_PRB2020} in the context of global quantum many-body scars. Here, we emphasize that the motif can be more generally used as a barrier to prevent heat flow between different regions irrespective of the Hamiltonian away from this barrier.) 

We simulate the Heisenberg Hamiltonian,
$H = \sum_{i,j} J_{ij} {\bf S_i \cdot S_j}$ on the lattice topology in Fig.~\ref{fig:systems}(b): $J_{ij}=1$ for all connected links except the shared common bond [bond (5,6)] which has the strength of $J_{ij}=2$. Each triangle [comprising sites $4,5,6$ and $5,6,7$ in Fig.~\ref{fig:systems}(b)] has two degenerate ``dimer covered" eigenstates, a consequence of the Hamiltonian for each triangle being proportionate to $({{\bf S}_i+ {\bf S}_j+{\bf S}_k})^2$. Much in the spirit of the Majumdar-Ghosh Hamiltonian~\cite{MG-model} (and other projector constructions~\cite{AKLT, Changlani_2018,Pal_2021}), the simultaneous eigenstates of both pasted triangles are simple -- they have a singlet on sites 5 and 6 and arbitrary spin states at the other vertices not part of this bond (i.e., sites 4 and 7). This property is ideal -- the time evolved state in the remainder of regions $A$ and $B$ has no bearing on the singlet, it remains frozen in time. This stems from a local integral of motion for total angular momentum of the spins, $Q=(S_5+S_6)^2$ -- the collective dynamics is nonergodic ($Q=0$) or ergodic ($Q=2$)~\cite{McClarty_PRB2020}.

We demonstrate our assertions in Fig.~\ref{fig:results-motif} using exact diagonalization. We prepare the ten-site system in the initial state $|\Psi \rangle = | \uparrow \uparrow \uparrow \uparrow \rangle \otimes |\psi_{5,6}\rangle \otimes |\text{ran}\rangle$ (i.e. sites 1-4 are in a ferromagnetic state, sites 5 and 6 are prepared in state $|\psi_{5,6}\rangle$ and sites 7-10 in a random state denoted by $|\text{ran}\rangle$). Note that the ferromagnetic state is the exact local eigenstate of the Heisenberg Hamiltonian in region $A$, however, the coupling to region $AB$ means it will in general (but not always) decohere under time evolution.

When $|\psi_{5,6}\rangle$ is prepared in a singlet state, we observe in Fig.~\ref{fig:results-motif}(a) that the half cut EE [dividing the bond (5,6)] is constant, at a value of one, as expected. In addition, Fig.~\ref{fig:results-motif}(b) shows that neither the total energy associated with the left nor right regions changes (i.e. there is no exchange of energy). Here, we define the energy for the left region as $\langle E_L\rangle=\langle \Psi(t)| \sum_{i,j}\mathbf{S}_i\cdot \mathbf{S}_j|\Psi(t)\rangle$, where $(i,j)=(1,2), (2,3), (3,4), (4,5), (4,6), (5,6)$. These are the bonds connecting the first four spins and the left triangle formed by spins $4$, $5$, and $6$. Similar, $(i,j)=(5,6), (5,7), (6,7), (7,8), (8,9), (9,10)$ are used to define the energy for the right region, $\langle E_R\rangle$. The ferromagnet in the left portion of the chain remains intact, owing to the fact that it was a local eigenstate of $A$ to begin with. This is verified in Fig.~\ref{fig:results-motif}(c), the plot of the local magnetization ($\langle S^{z}_j (t)\rangle$ for each site $j$). The sites of region $B$ being initially prepared in a random state evolve in time and eventually appear to thermalize. We also plot the von Neumann entropy for each site, and find that it remains low in $A$ but large in $B$ [see Fig.~\ref{fig:results-motif}(e)].

In contrast, when $|\psi_{5,6}\rangle$ is prepared in a triplet state $|T\rangle=(|\uparrow\downarrow\rangle+|\downarrow\uparrow\rangle)/\sqrt{2}$, the half-cut EE quickly grows and saturates. The energy exchange between the left and right regions is also rapid. Though the bond energies and entropy are very distinct to begin with [as can be seen from the inset of Fig.~\ref{fig:results-motif}(f)], they assume a much more uniform value as time progresses. Qualitatively similar behavior is seen (for all the observables reported here) when $|\psi_{5,6}\rangle$ is prepared in a random linear superposition of $|\uparrow\uparrow\rangle$, $|\uparrow\downarrow\rangle$, $|\downarrow\uparrow\rangle$, and $|\downarrow\downarrow\rangle$.

\textit{Conclusion--} To conclude, we have provided a framework for how thermalization in quantum systems can be arrested in a controlled way. Our results stem from a combination of Hilbert space fragmentation and local symmetry~\cite{Sala2020, Khemani2020, Moudgalya2021, Buca-arxiv, Buca_PRL2022}. We provided a proof of principle of our analytic assertions that were verified numerically in systems of few spins. All qualitative conclusions presented here were also found to hold for systems with a few additional spins (not shown here). Using a tilted field Ising model, our first exploration demonstrated how energy flow can be arrested when the interaction between the subsystems satisfy a commutation condition. The dynamics of the local region is decoupled from the rest of the system. Second, for the case of noncommuting subsystems, we found that it is possible to arrest thermalization with the help of local motifs which have eigenstates that do not evolve and hence do not affect the evolution in other regions. In the example we presented, when two sites in the boundary region are initially entangled in a singlet configuration, there is no flow of energy between the two subsystems. When the initial state is changed, the two subsystems exchange energy. This is precisely the function of a thermal switch. 

We anticipate the realization of our ideas on multiple promising platforms - cold atoms~\cite{Kimble2016}, Rydberg atoms~\cite{Zoller2010, Lahaye2015, Bloch-PRX2017, Lahaye2020}, and circuit QED~\cite{Petersson2012, Wallraff2014, Wallraff2015}. More generally, our results suggest that manipulating a small number of spins and/or the interactions between them can completely change the thermalizing properties of the entire system. We imagine the use of these concepts
to effectively disconnect or ``insulate" systems in a controlled way. Finally, we also envision the generalization to higher dimensional systems, for example, by introducing an arbitrary number of small local insulating puddles. 

\begin{acknowledgments} 
H.J.C. thanks A. Pal, B. Mukherjee, M. Szyniszewski, P. Sharma and R. Melendrez for useful discussions and for collaborations on related work. We also thank Berislav Bu\v{c}a for useful comments. We thank Florida State University and the National High Magnetic Field Laboratory for support. The National High Magnetic Field Laboratory is supported by the National Science
Foundation through NSF/DMR-1644779 and DMR-2128556 and the State of Florida. The work by KKWM was supported by the Dirac postdoctoral fellowship. H.J.C. acknowledges support from the National Science Foundation Grant No. DMR-2046570.
\end{acknowledgments}

\end{document}